\documentclass[12pt]{spieman}  
\usepackage{amsmath,amsfonts,amssymb}
\usepackage{graphicx}
\usepackage{setspace}
\usepackage{tocloft}
\usepackage{float}
\usepackage{subfig}
\usepackage{color}
\usepackage{multirow}
\usepackage{caption}

\usepackage[table,xcdraw]{xcolor}

\DeclareMathOperator*{\minimize}{minimize}

\title{Convolutional Neural Networks Based Texture Modeling For AV1}
\author[a]{Di Chen}
\author[a]{Chichen Fu}
\author[b]{Zoe Liu} 
\author[a]{Fengqing Zhu}
\affil[a]{Purdue University, School of Electrical and Computer Engineering, 465 Northwestern Avenue, West Lafayette, Indiana 47907, USA}
\affil[b]{Visionular Inc, 4290 Briarwood Way, Palo Alto, California 94306, USA}

\cftpagenumbersoff{figure}
\cftpagenumbersoff{table} 
\begin{document} 
\maketitle

\begin{abstract}
Modern video codecs including the newly developed AOMedia Video 1 (AV1) utilize hybrid coding techniques to remove spatial and temporal redundancy. However, efficient exploitation of statistical dependencies measured by a mean squared error (MSE) does not always produce the best psychovisual result. One interesting approach is to only encode visually relevant information and use a different coding method for ``perceptually insignificant" regions in the frame, which can lead to substantial data rate reductions while maintaining visual quality. In this paper, we introduce a texture analyzer before encoding the input sequences to identify ``perceptually insignificant" regions in the frame using convolutional neural networks. We designed and developed a new scheme that integrate the texture analyzer into the codec that can largely reduce the temporal flickering artifact for codec with hierarchical coding structure. The proposed method is implemented in AV1 codec by introducing a new coding tool called texture mode, where texture mode is a special inter mode treated at the encoder, that if texture mode is selected, no inter prediction is performed for the identified texture regions. Instead, displacement of the entire region is modeled by just one set of motion parameters. Therefore, only the model parameters are transmitted to the decoder for reconstructing the texture regions. Non-texture regions in the frame are coded conventionally. 
We show that for many standard test sets, the proposed method achieved significant data rate reductions with satisfying visual quality.

\end{abstract}

{\noindent \footnotesize\textbf{*}Di Chen,  \linkable{chen1939@purdue.edu} }
{\noindent \footnotesize\textbf{*}Chichen Fu,  \linkable{fu26@purdue.edu} }
{\noindent \footnotesize\textbf{*}Zoe Liu,  \linkable{zoeliu@gmail.com} }
{\noindent \footnotesize\textbf{*}Fengqing Zhu,  \linkable{zhu0@ecn.purdue.edu} }

\begin{spacing}{2}   

\section{Introduction}
\label{sect:intro}  

Modern video codecs utilize hybrid coding techniques consisting of 2D transforms and motion compensation techniques to remove spatial and temporal redundancy. However, efficient exploitation of statistical dependencies measured by a mean squared error (MSE) does not always produce the best psychovisual result. One approach is to only encode areas of a video frame that are ``perceptually significant" using hybrid coding techniques. The ``perceptually insignificant" regions will be encoded using statistical models of the pixels.  By ``perceptually insignificant" pixels we mean regions in a frame that an observer will not notice any difference without observing the original video sequence.  The perceptually insignificant pixel blocks usually are ``noise-like" textures and are generally costly to encode using hybrid coding methods. In this paper, we define texture area as  ``perceptually insignificant" pixels in a frame with respect to the Human Vision System while ``none-texture" area refers to all other pixels in the frame. The encoder instead fits a statistical model to perceptually insignificant pixels in the frame and transmits the model parameters to the decoder as side information. The decoder uses the model to reconstruct the pixels. This is referred to as “analysis/synthesis” coding.  Since the parameters of the statistical model generally can be represented by far fewer bits, the data rate can be significantly reduced.  

The problem with using “analysis/synthesis” approach is that if perceptually insignificant pixel blocks in each frame are encoded separately using the model, the areas that have been reconstructed with the model may be visible when the video is displayed because the model does not take into account the temporal variation in the scene. This requires spatial and temporal segmentation and labeling methods and models. In our previous work \cite{bosch2011}, we developed a feature based texture analyzer to identify perceptually insignificant regions in the frame. At the encoder, instead of performing inter-frame prediction to reconstruct these regions, displacement of the entire texture region is modeled by a set of motion parameters. We have shown that data rate reductions of 5-20\% can be achieved with this approach \cite{bosch2011} using a hand-crafted feature based texture analyzer when implemented using H.264.

There are other works that use perceptual based approaches to improve the performance of video codec. In \cite{Balle2011}, both static and dynamic texture models are investigated for image and video coding, respectively. A texture analysis-synthesis loop is developed to identify ``rigid" or ``non-rigid" texture regions which are then synthesized at the decoder using corresponding side information. This approach only applies to Gaussian texture. In \cite{fan2011}, a perspective motion model is employed to warp static textures. Texture regions are segmented using features derived from wavelet transform and further classified based on their spatial and temporal features. It relies on the proper choices of a set of parameters to achieve accurate texture segmentation. The texture synthesizer in \cite{karam2015} does not use motion model but instead uses pixels from the above and left macroblocks to synthesize static texture. In practice, this requires the current block to find similar contexts from the above and left area which may not be true for all encoding blocks.

While the hand-crafted feature based texture analyzer requires a set of manually chosen parameters to achieve accurate texture segmentation for different videos, deep learning based methods usually do not require such parameter tuning for inference since these parameters are learned during the training step. In our previous work \cite{fu2018}, we proposed a block-based texture segmentation method to extract texture regions in a video frame using convolutional neural networks and only encode areas of a video frame that are ``perceptually significant." 
 

In this paper, we extend the idea in \cite{fu2018} by refining the texture mask and integrating learning based texture analysis into the modern hybrid codec workflow.

The problem with using the texture analyzer alone to encode the texture region in the video is that if each frame is encoded separately, areas that are reconstructed using the texture models will be obvious when the video is displayed. This then requires the textures to be modeled both spatially and temporally. 
We propose a new scheme that reconstructs the texture region differently for the bidirectional predicted frames (B-frames) and forward predicted frames (P-frames) which largely reduces the temporal visual artifacts. 
The proposed method is implemented using AV1 codec\cite{av1-joshi2017,av1-liu2017,chen2018,fu2018}, which is an open-source and royalty-free video codec developed by The Alliance for Open Media (AOM) \cite{AOM}. We introduce a new coding mode for AV1 called the ``texture mode". The ``texture mode" is completely an encoder side option, which in essence skips the coding of the block entirely through leveraging the use of global motions provided by the AV1 baseline. Specifically, the texture mode uses a modified version of the global motion coding tool in the AV1 codec \cite{global_motion} to ensure temporal consistency of the texture regions between frames. Based on the selection of coding structures and choices of reference frames, we investigate three different implementations of the ``texture mode" in terms of data rate savings and perceived quality. Experimental results validate the efficacy of the texture mode with a consistent coding gain compared to the AV1 baseline over a variety of video test sets given a fixed perceptual quality level. We also conducted a subjective visual quality test to validate the proposed approach. 

The main contributions of this paper are summarized as follows:
\begin{itemize}
\item[1] We extended our work in \cite{fu2018} by improving the Convolutional Neural Network architecture which identifies the ``perceptually insignificant" region before encoding the video sequences using conventional video codec. A series of post-processing steps are added to refine the texture mask, which improves the texture identification accuracy.
\item[2] We developed a new scheme that reconstructs the texture region differently for the B-frames and the P-frames which largely reduces the temporal visual artifacts. 
\item[3] Our proposed method is a pioneering work that integrates learning-based texture analysis and reconstruction approach with modern video codec for enhancing video compression performance.
\end{itemize}

The paper is organized as follows: Section 2 describes the CNN-based texture analyzer and the proposed ``texture mode" to integrate the texture analyzer into the AV1 video codec. Section 3 shows the experimental results for the texture analyzer and coding performance, as well as a detailed discussion of our observation. Conclusion is provided in Section 4.
\section{Texture-Base Video Coding}
\begin{figure}[ht]
\includegraphics[width=1\columnwidth]{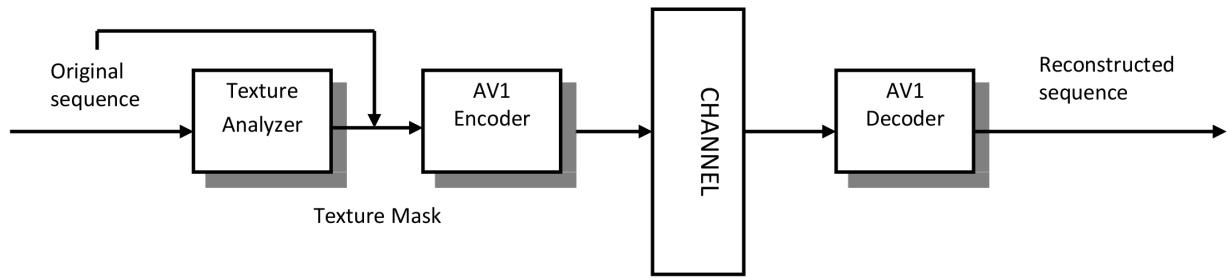}
\caption{Overview of texture-based video coding}
\label{fig:blockdia}
\end{figure}

The general scheme for video coding using texture analysis and reconstruction is illustrated in Fig. \ref{fig:blockdia}. The texture analyzer identifies the texture regions in a frame. We use a classification convolutional neural network to label each block in a frame as textures or non-texture and generate a block-based texture mask for each frame. The texture mask and the original frame are fed into the AV1 video codec to enable the texture mode where the identified texture regions skip the encoding process. The texture region is reconstructed by warping texture region in a reference frame to the current frame. A modified version of the global motion tool \cite{global_motion} in AV1 is used to obtain motion estimation and reconstructs the texture region without sending residues for the identified texture region.

\subsection{Texture Analysis Using CNN}
\label{ssec:texcnn}
\begin{figure}[ht]
\includegraphics[width=0.95\columnwidth]{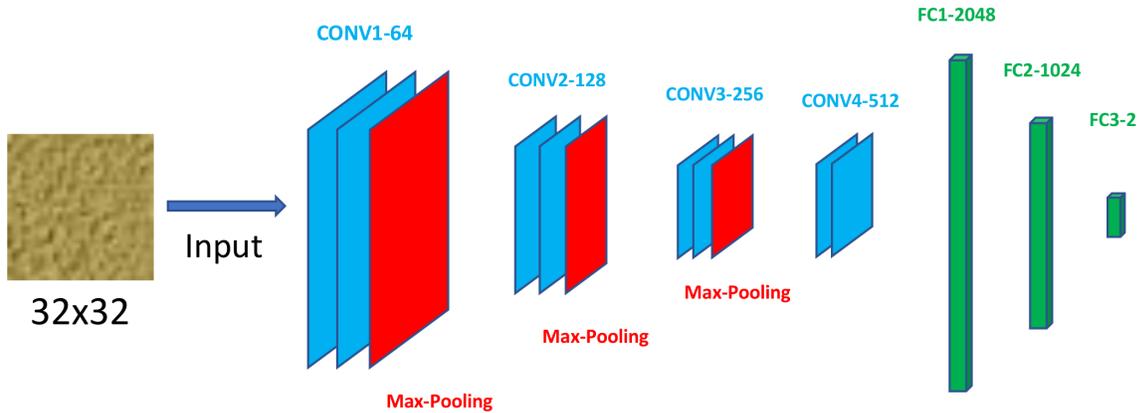}
\caption{CNN architecture for block-based texture classification}
\label{fig:archit}
\end{figure}

Based on the block-based segmentation method introduced in \cite{fu2018}, we identify texture regions in each frame which are potential candidates to use the texture mode in AV1. We designed a classification convolutional neural network inspired by the VGG network architecture \cite{Simonyan2014} to label a block as texture or non-texture (Fig. \ref{fig:archit}). The input of our network is a $32\times32$ color image block. The output is the probability that the image block contains texture or non-texture.

\begin{figure}[ht]
\includegraphics[width=1\columnwidth]{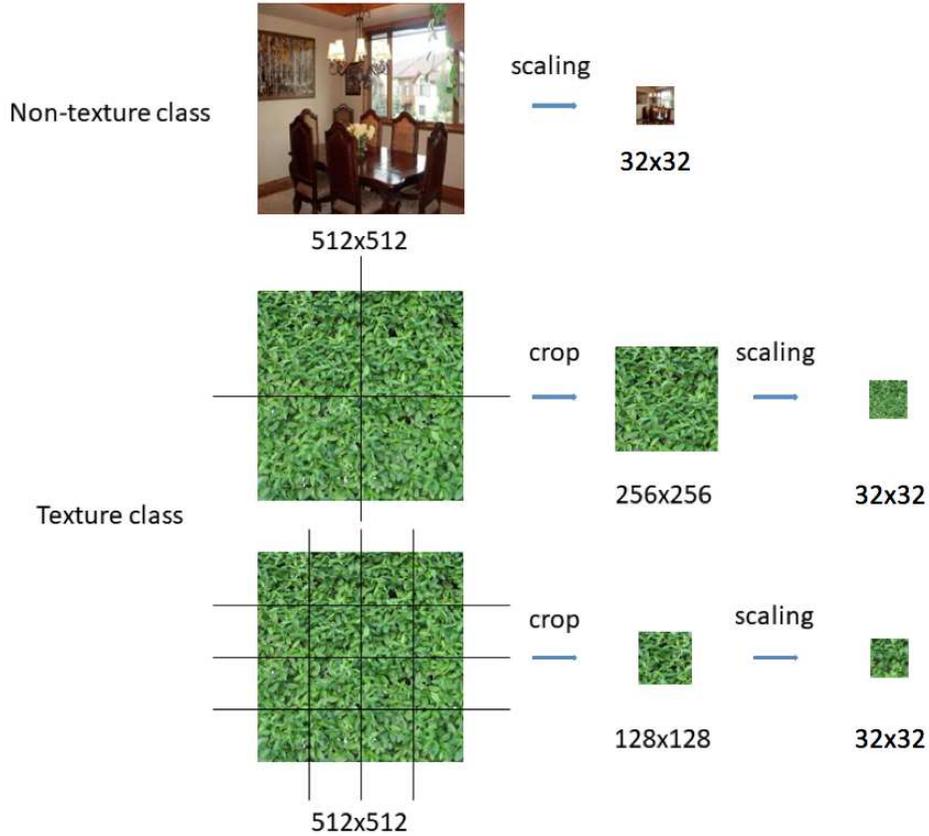}
\caption{Training data preparation}
\label{fig:datapro}
\end{figure}

In our previous work\cite{fu2018}, image patches of size $16\times16$ with texture and non-texture labels are used to train the network. Since the image patch size is small, a block within a small moving object can also be detected as a texture block. In this paper, the training image patch size is increased to $32\times32$ to minimize encoding artifacts. Texture and non-texture images are obtained from the Salzburg Texture Image Database (STex) \cite{stex} and Places365 \cite{place365}. STex contains images with single texture type and images in Place365 are nature scenes with multiple objects. To create multi-resolution training samples for texture classes, images from STex are cropped from $512\times512$ into $256\times256$ and $128\times128$, respectively, followed by downsampling to $32\times32$. For non-texture class, images from Place365 are directly downsampled to $32\times32$ image patches to create non-texture examples that contain multiple objects and structural features. The data preparation is shown in Fig. \ref{fig:datapro}.

This method was implemented in Torch \cite{torch}. A stochastic gradient descent (SGD) with momentum is used to train our network. A learning rate of 0.01, a momentum of 0.9 and weight decay of 0.00005 were used in our training. A set of training data with 1,740 texture examples and 36,148 non-texture examples were used to train our network. A binary cross entropy loss was used as the loss function. Since our training set is highly unbalanced, the weights of each class in the binary cross entropy loss function were set proportion to the inverse of the class frequency. A total of 100 epochs were trained using a mini batch size of 512 on one NVIDIA GTX TITAN GPU.

After training the CNN, texture segmentation is performed on each test video frame. Each frame is divided into $32\times32$ adjacent non-overlapping blocks. Each block in the video frames is classified as either texture or non-texture. The segmentation mask for each frame is formed by grouping the classified blocks in the frame.

\begin{figure}[ht]
\includegraphics[width=1\columnwidth]{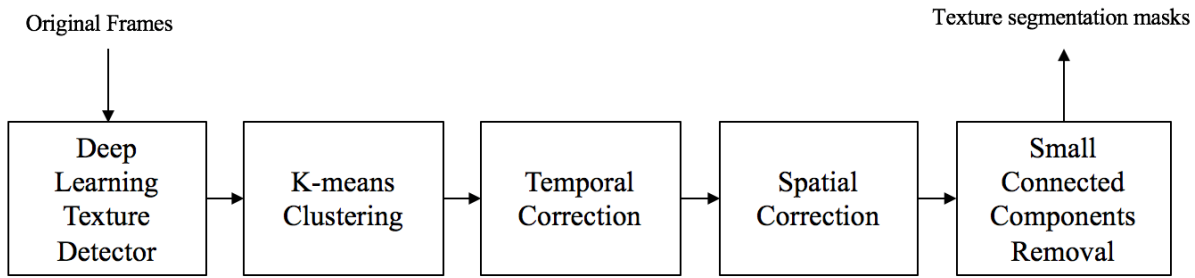}
\caption{Flowchart of texture analyzer}
\label{fig:postpro}
\end{figure}

\subsection{Texture Segmentation Mask Refinement}
In order to fit the texture segmentation mask in the AV1 codec and minimize the artifacts produced by encoding, a series of processes are used.  As shown in Fig. \ref{fig:postpro}, an adaptive K-means clustering \cite{chen1998} is used to group different types of texture in the texture segmentation mask. A temporal and spatial correction is then used to maintain the consistency of texture segmentation mask throughout the entire video sequence.  The temporal correction uses a majority voting of three consecutive frames to determine whether a block in the current frame should be labeled as a texture block. The spatial correction uses 4-connectivity neighborhood voting to fill holes in the texture segmentation mask. Finally, a small connected labeling marks the connected components that are less than 5 blocks as non-texture blocks. 

\subsection{A New AV1 Coding Tool - Texture Mode}
In this section, we describe how we modified the AV1 codec by introducing a texture mode to encode the identified texture blocks in a video frame.

\subsubsection{Texture Mode Encoder Design}
\begin{figure}[t]
\begin{center}
\noindent
  \includegraphics[width=0.7\columnwidth]{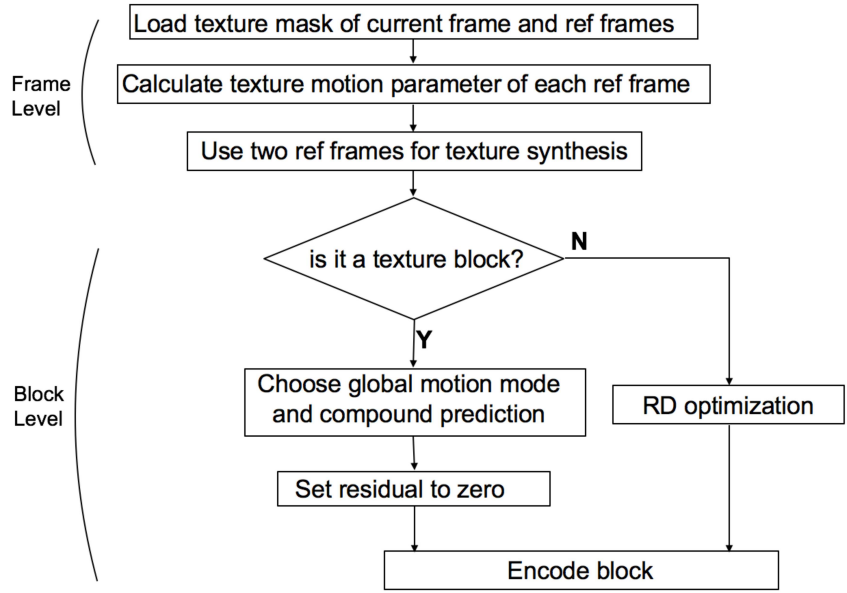}
  \caption{Texture mode encoder implementation}
  \label{Fig1Label}
\end{center}
\end{figure}

The texture analyzer is integrated into the AV1 encoder as illustrated in Fig. \ref{Fig1Label}. At the encoder, for each frame that contains texture area, we first fetch the texture masks for the current frame and the selected reference frames from the texture analyzer. 
Based on the texture region in the current frame, a set of texture motion parameter that represents the global motion of the texture area is estimated for each reference frame. For each block larger than $32\times32$, we use a two-step method to check if a block is a texture block as described in section \ref{sect:texb_decision}.
A texture block is reconstructed using global motion parameters and no motion compensation residuals will be coded and transmitted for the texture block. We call this new coding paradigm the texture mode. At the decoder, since there is no syntax change to the AV1 bitstream, the bitstream is decoded the same as AV1 baseline.

\begin{table*}[ht]
\scriptsize
\begin{center}
\caption{Configuration of different texture mode implementation}
\label{table:config}
\begin{tabular}{|c|c|c|}
\hline
tex-all                                                     & tex-sp                                        & tex-cp                                   \\ \hline
\multicolumn{3}{|c|}{Disable texture mode for GOLDEN / ALTREF frame}                                                                                     \\ \hline
Original GF group interval (4-16)                           & \multicolumn{2}{c|}{Fixed 8 GF group interval}                                             \\ \hline
single-layer coding structure                               & \multicolumn{2}{c|}{multi-layer coding structure}                                          \\ \hline
Use texture mode for all frames except GOLDEN /ALTREF frame & \multicolumn{2}{c|}{Use texture mode for every other frame (frame1,3,5,7 in the GF group)} \\ \hline
Use single-prediction (forward or backward)                 & Use single forward prediction                   & Use compound prediction                  \\ \hline
\end{tabular}
\end{center}
\end{table*}

\begin{figure}
\begin{minipage}[htp]{0.05\textwidth}\raggedleft
(a)  \\
\vspace{4.5cm}
(b)  \\
\vspace{4.5cm}
(c)  \\
\end{minipage}
\begin{minipage}[htp]{14cm}
\includegraphics[width=\linewidth]{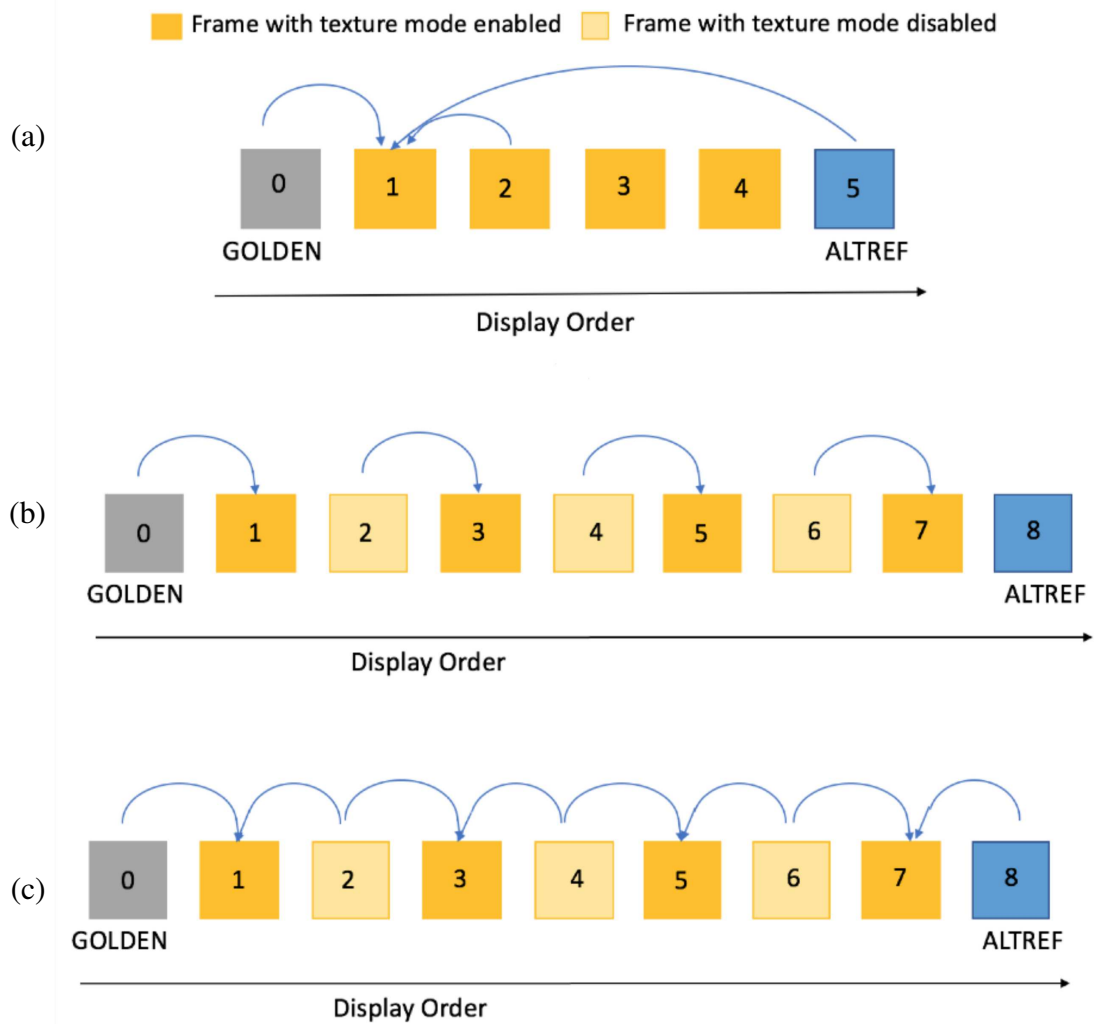} 
\end{minipage}
\captionof{figure}[Coding structure of texture mode]{Coding structure of texture mode: (a) GF group coding structure using tex-all configuration. (b) GF group coding structure using tex-sp configuration. (c) GF group coding structure using tex-cp configuration.}
  \label{gf_config}
\end{figure}

In general, a texture block in the current frame is reconstructed by warping the texture block from the reference frame towards the current frame. We use a modified version of the global motion coding tool \cite{global_motion} in the AV1 codec to perform block warping as described in Section \ref{sssec:motion}. 
Based on the selection of coding structures and choices of reference frames for texture reconstruction, we investigated three different implementations, namely tex-all, tex-sp, and tex-cp of the texture mode in terms of data rate savings and perceived quality. Configuration of the three implementations are described in Table \ref{table:config} and can be visualized in Fig. \ref{gf_config}. Encoders of AOM/AV1 codec consider an input video sequence as a succession of frames grouped in Golden-Frame (GF) groups. GF groups may have 4-16 frames with the first frame being the GOLDEN frame and the last frame being ALTREF frame. For tex-sp and tex-cp, a multi-layer hierarchical coding structure \cite{multilayer} is used for each GF group.

\noindent

The tex-all implementation has the best data rate savings since the number of frames with texture mode enabled is approximately twice as many as the other two implementations. However, we noticed some visual artifacts in the reconstructed videos in several test sequences due to the accumulated errors from warping displacement. If a block contains both texture and non-texture region and is classified as a texture block, there will be visual artifacts in the non-texture region. These artifacts can be passed on to other frames that use this frame as a reference frame, and causing the artifacts to propagate. The artifacts are most prominent in videos with high motion or complex global motion.

The tex-sp implementation solves the accumulation error by only enabling texture mode for every B frame and disabling texture mode for every I and P frame. It only uses the immediate previous frame as the reference frame for texture warping to estimate a more accurate global motion model. As a result, the data rate savings are reduced to approximately half the data rate savings of the tex-all configuration. Some flickering artifacts can still be observed between frames for some test videos.

The tex-cp further reduces the flickering artifacts by using compound prediction from the previous frame and the next frame. The data rate savings are only slightly lower than that of tex-sp. The improvement in visual quality is most prominent in low-mid resolution videos. In practice, we use tex-cp as the configuration for the texture mode which ensures good visual quality.

\subsubsection{Texture Motion Parameters}
\label{sssec:motion}
The global motion coding tool \cite{global_motion} in AV1 is used primarily to handle camera motion. A motion model is explicitly conveyed at the frame level for the motion between a current frame and any one or more of its reference frames. The motion model can be applied to any block in the current frame to generate a predictor. The affine transformation is selected as the motion model. The motion model is estimated using a FAST feature \cite{FAST} matching scheme followed by a robust model fitting using RANSAC \cite{RANSAC}. The estimated global motion parameter is added to the compressed header of each inter-frame. 

Since the motion model parameters of the global motion coding tool in AV1 is estimated at the frame level between the current frame and the reference frame, these parameters may not accurately reflect the motion model for the texture regions within a frame. We modified the global motion tool to design a new set of motion modal parameters, called texture motion parameters. The texture motion parameters are estimated based on the texture region of the current frame and the reference frame using the same feature extraction and model fitting method as in the global motion coding tool. A more accurate motion model for texture region may reduce the artifacts on the block edges between the texture blocks and non-texture blocks in the reconstructed video. In order to keep the syntax of AV1 bitstream consistent, the texture motion parameters are sent to the decoder in the compressed header of the inter frames by replacing the global motion parameters of the reference frames. 



\subsubsection{Texture Block Decision}
\label{sect:texb_decision}
The minimum size of a texture block is $32 \times 32$ as described in Section \ref{ssec:texcnn}. For all blocks larger than or equal to $32 \times 32$, we use a two-step approach to check if a block should be encoded using the novel texture mode scheme we proposed to AV1. First, we overlap the texture mask generated by the texture analyzer and the current frame to check if the entire block is inside the texture region of the current frame. We also need to ensure that the pixels used for texture reconstruction in the reference frames are within the texture regions identified by the texture analyzer. In order to maintain temporal consistency of the texture regions, in the second step, we warp the blocks inside the texture region towards the two reference frames, i.e., the previous frame and the next frame in the tex-cp configuration. If the two warped blocks are within the texture regions of both corresponding reference frames, the block is considered a texture block and texture mode is enabled. 

\subsubsection{Block Splitting Decision}
Like most modern video coding standards, the baseline AV1 codec uses rate-distortion (RD) optimization to make block splitting decision and choose the best coding units. The problem of finding the block splitting decision that minimizes the distortion between the original block and its reconstruction subject to a constrained rate can be described as
\begin{equation}
\displaystyle{\minimize_{p\in P_{k}} D_{k}(p)+\lambda R_{k}(p)}
\end{equation}
where $R_{k}(p)$ represents the number of bits that are required for signaling the block splitting and prediction decision $p$ and the actual bit cost by an entropy coder in the bitstream. $D_{k}(p)$ represents the distortion measure. $\lambda \geq 0$ denotes the Lagrange multiplier.

In our proposed method, however, the position of the texture regions inside a block has higher priority than the RD values of different block splitting methods for this block. If the block is a texture block, we do not further split it into smaller sub-blocks. If the block contains no texture region, RD optimization is performed for block partitioning and mode decision. If part of a block contains texture region, we split it into sub-blocks regardless of the RD value as long as the size of the sub-blocks is equal or larger than $32\times32$. In general, there is no block that is part texture and part non-texture. The use of texture mode also largely increases the encoding speed, since no RD optimization is performed for a texture block which reduces the need for different prediction modes, reference frames selection, and the block splitting recursion.

\subsubsection{Texture Reconstruction}
We use AV1 codec's global motion tool and compound prediction to reconstruct texture for texture blocks at the decoder. The previous frame and the next frame of the current frame are chosen to be the reference frames for the texture block reconstruction. The texture regions in the two reference frames are warped towards the texture blocks in the current frame using the corresponding texture motion parameters. We used compound prediction to reconstruct the texture block from the two reference frames. As discussed earlier, the use of compound prediction for texture blocks reduces flickering artifacts between frames. The residual of the texture blocks is set to zero. Since all the texture blocks in one frame use the same reference frames and the same set of motion parameters, there is no displacement on the block edges of the texture blocks within the texture region.  
\section{Experimental Results}

\subsection{Texture Analysis}
\begin{figure}[ht]
\centering
\includegraphics[width=0.6\columnwidth]{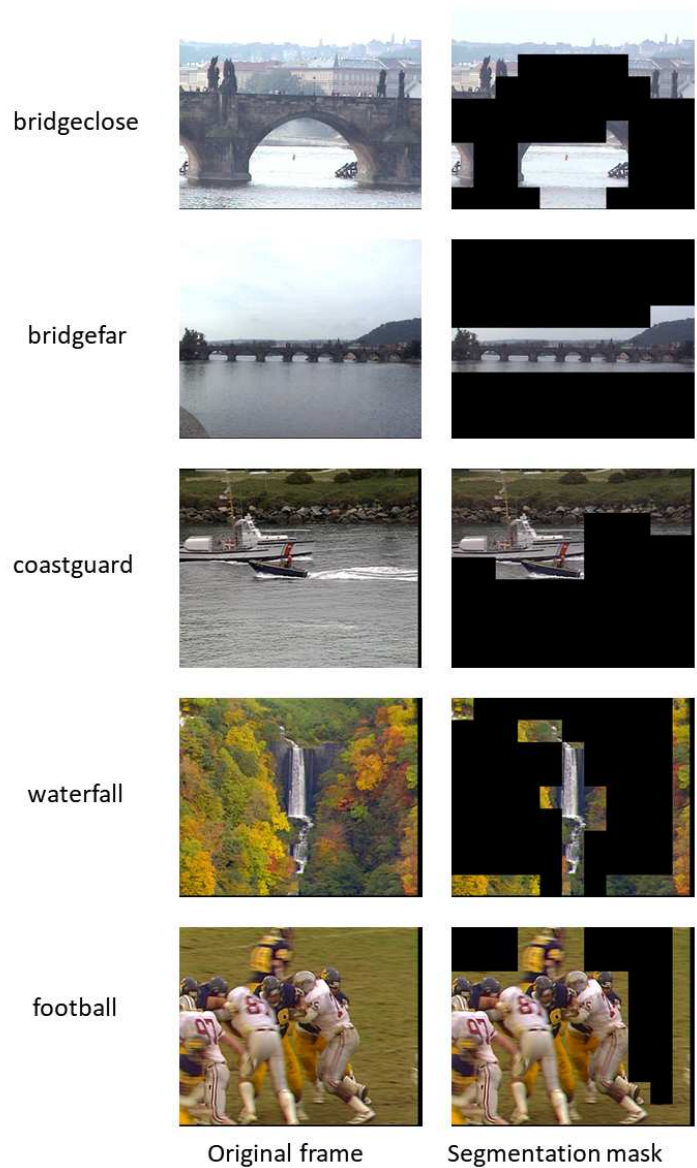}
\caption{Texture segmentation examples}
\label{fig:tex_res}
\end{figure}

Nine representative video sequences contain different texture types were tested using the CNN based texture analyzer. Sample texture segmentation results are shown in Fig. \ref{fig:tex_res}. Our texture analyzer can successfully identify and locate most texture regions. Currently, the texture analyzer uses a block-based texture classification method with fixed block size. Texture within smaller blocks is not included in the segmentation mask. For example, in the sequence bridgefar, some parts of the water, as well as the sky, are not included in the texture mask because they cannot cover a $32\times 32$ block. Small non-texture parts can also be included in the segmentation mask when the majority of the $32\times 32$ block is in a texture region, such as the bowing of the white boat in the sequence coastguard.

\subsection{Coding Performance}
To evaluate the performance of the proposed texture-based method, data rate savings at four quantization levels (QP = 16, 24, 32, 40) are calculated for each test sequence using the tex-cp configuration and compared to the AV1 baseline. The AV1 baseline is the original codec with a fixed group interval of eight frames and using hierarchical multilayer coding structure \cite{multilayer}. The data rate is computed by dividing the output WebM file size by the number of frames. The WebM file is the output bitstream from the AV1 encoder. Results for several test videos are shown in Table \ref{gain}. The test videos include large texture areas. We also include the average percentage of pixels that uses the texture mode in a frame in the table.   
%
\begin{table}[t]
\begin{center}
  \caption{Data rate savings for different QP level}
  \label{gain}

\small\addtolength{\tabcolsep}{-0.3pt}
\begin{tabular}{|c|c|c|c|c|c|c|c|}
\hline
\multirow{2}{*}{Video} & \multirow{2}{*}{Resolution} & \multirow{2}{*}{\begin{tabular}[c]{@{}c@{}}Number of\\ Frames\end{tabular}} & \multicolumn{4}{c|}{Data Rate Saving (\%)} & \multirow{2}{*}{\begin{tabular}[c]{@{}c@{}}Texture \\ Region (\%)\end{tabular}} \\ \cline{4-7}
                       &                             &                                                                             & QP=16     & QP=24    & QP=32    & QP=40    &                                                                            \\ \hline
bridgeclose            & cif                         & 300                                                                         & -13.77    & -9.78    & -3.20    & 1.67     & 33.12                                                                      \\ \hline
bridgefar              & cif                         & 300                                                                         & -8.80     & -5.90    & -4.44    & -4.38    & 21.19                                                                      \\ \hline
coastguard             & cif                         & 300                                                                         & -7.80     & -6.99    & -4.70    & -1.90    & 37.42                                                                      \\ \hline
flower                 & cif                         & 150                                                                         & -10.55    & -8.66    & -5.96    & -3.38    & 58.00                                                                      \\ \hline
football               & cif                         & 150                                                                         & -0.35     & 0.02     & 0.01     & 0.02     & 10.32                                                                      \\ \hline
waterfall              & cif                         & 150                                                                         & -4.63     & -3.96    & 0.33     & 3.74     & 61.22                                                                      \\ \hline
netflix\_aerial        & 512x270                     & 300                                                                         & -8.59     & -2.15    & 0.68     & 4.59     & 29.94                                                                      \\ \hline
netflix\_rollercoaster & 512x270                     & 300                                                                         & -3.44     & -2.39    & -1.02    & 0.41     & 35.55                                                                      \\ \hline
intotree               & 1280x720                    & 300                                                                         & -5.32     & -4.23    & -1.99    & 2.83     & 43.25                                                                      \\ \hline
tractor                & 1280x270                    & 300                                                                         & -3.32     & -2.25    & -1.68    & 0.30     & 20.17                                                                      \\ \hline
crowd\_run                & 1920x1080                    & 300                                                                         & -0.16     & -0.10    & -0.05    & 0.10     & 9.45                                                                      \\ \hline
\end{tabular}
\end{center}
\end{table}

As shown in Table \ref{gain}, at low QP, most of the videos show large data rate savings. However, as the QP increases, the data rate savings decreases. At high QP, texture-based method tends to have worse coding performance than AV1 baseline for some test videos, such as football, waterfall and netflix\_aerial. This is because at high QP, many non-texture blocks also have zero residual and our method requires a few extra bits for the texture motion parameters and for using two reference frames in compound prediction. 

\subsection{Result For Subjective Visual Quality Test}

\begin{table}[]
\small\addtolength{\tabcolsep}{-0.3pt}
\centering
\caption{Result for subjective visual quality test}
\label{table:subjective}
\begin{tabular}{|c|c|c|c|}
\hline
Video                  & {\color[HTML]{000000} \begin{tabular}[c]{@{}c@{}}Quality of reconstructed \\ video from tex-cp \\ is \textbf{better} than \\ the original codec\end{tabular}} & \begin{tabular}[c]{@{}c@{}}Quality of reconstructed \\ video from tex-cp\\  is \textbf{equal} to \\ the original codec\end{tabular} & \begin{tabular}[c]{@{}c@{}}Quality of reconstructed \\ video from tex-cp\\  is \textbf{worse} than \\ the original codec\end{tabular} \\ \hline
bridgeclose            & 15\%                                                                                                                                                 & 60\%                                                                                                                       & 25\%                                                                                                                         \\ \hline
bridgefar              & 10\%                                                                                                                                                 & 65\%                                                                                                                       & 25\%                                                                                                                         \\ \hline
coastguard             & 40\%                                                                                                                                                 & 40\%                                                                                                                       & 20\%                                                                                                                         \\ \hline
flower                 & 20\%                                                                                                                                                 & 55\%                                                                                                                       & 25\%                                                                                                                         \\ \hline
football               & 0\%                                                                                                                                                  & 65\%                                                                                                                       & 35\%                                                                                                                         \\ \hline
waterfall              & 25\%                                                                                                                                                 & 60\%                                                                                                                       & 15\%                                                                                                                         \\ \hline
netflix\_aerial        & 20\%                                                                                                                                                 & 75\%                                                                                                                       & 5\%                                                                                                                          \\ \hline
netflix\_rollercoaster & 25\%                                                                                                                                                 & 60\%                                                                                                                       & 15\%                                                                                                                         \\ \hline
intotree               & 15\%                                                                                                                                                 & 55\%                                                                                                                       & 30\%                                                                                                                         \\ \hline
\end{tabular}
\end{table}

Established quality assessment measures, such as the PSNR cannot be applied to evaluate our method because there can be a large pixel-wise difference in the expected MSE between texture regions reconstructed using motion compensation versus using the proposed method. However, an observer will not notice any difference without observing the original video sequence. A similar argument holds for the structural similarity index (SSIM) as it is based on the sample cross-covariance. Therefore, we performed a subjective visual quality study on 20 subjects. The study received Purdue Institutional Review Board approval under protocol \#1802020229. All subjects have signed the consent forms prior to participating in the research. 

In the study, each participant is asked to watch two versions of a test video. One is the reconstructed video using the original AOM/AV1 codec with QP=16. The other is the reconstructed video using our proposed method (tex-cp) with QP=16. Then they are asked to choose the video that has better visual quality. The subjects did not know which method is used to encoding the video. They were asked to choose among three options: the first video has better visual quality, the second video has better visual quality, or there is no difference between the two versions. The two versions of test video are shown one at a time and participants can watch the videos as many times as they would like to. The result of this study is summarized in Table \ref{table:subjective}. 

Results show that on average 59\% of the times participants cannot tell the difference between the reconstructed video by the original codec and the proposed method. 19\% of the times participants prefer the visual quality of the reconstructed video by the proposed method. 22\% of the time the visual quality of the reconstructed video using baseline AV1 is preferred. This indicates that the visual quality degradation due to our video coding models is minimal and acceptable. The texture region does not use residual for reconstruction. Although they have quality degradation with respect to PSNR, an observer will not notice any difference without observing the original video sequence. The main artifacts mainly come from the inaccurate block-based texture mask. For example, the football sequence has flickering artifacts on some of the frames which is noticeable to the participant. In the proposed texture analyzer, a block is considered a texture block if it contains a large percentage of texture area, although it may also contain a small amount of structural objects. In the football sequence, some parts of the player are also included in the texture mask and reconstructed using texture mode. Since the motion of the player is different from the motion of the texture area, i.e. the grass, there are noticeable flickering artifacts around those parts of the frame. This phenomenon also happens to some frames in the coastguard sequence. We plan to improve the texture analyzer in our future work by generating a more accurate segmentation mask of the texture area to reduce these artifacts.  Another visual artifact we noticed comes from the inaccurate motion model. For example, the intotree sequence has the most complex global motion among all the test sequences which may be better presented using the planar perspective motion model instead of the affine motion model used in the global motion model. 
\section{Conclusion}
In this paper, we proposed a new video coding paradigm that integrates texture modeling into a video codec. The texture modeling method uses deep learning based approaches to detect texture regions in a frame that is perceptually insignificant to the human visual system. We developed a scheme to integrate the texture analyzer into the video codec that reconstructs the texture region differently for B/P frames which largely reduces the temporal visual artifacts. The proposed method is implemented and tested using the AV1 codec by introducing a new coding mode - texture mode, to the AV1 encoder. The texture mode uses the multi-layer coding structure, a modified global motion tool and the compound prediction mode. Our results showed significant increase in terms of coding efficiency compared to the AV1 baseline for a representative set of video sequences containing large texture regions.


\acknowledgments 
This work is partially sponsored by Google Inc. Any opinions, findings, and conclusions or recommendations expressed in this material are those of the author(s) and do not necessarily reflect the views of Google Inc. Address all correspondence to Fengqing Zhu, zhu0@ecn.purdue.edu.

\bibliography{texture}   

\begin{thebibliography}{10}

\bibitem{bosch2011}
M.~Bosch, F.~Zhu, and E.~J. Delp, ``Segmentation{-}based video compression
  using texture and motion models,'' {\em IEEE Journal of Selected Topics in
  Signal Processing} {\bf 5}(7), 1366--1377  (2011).

\bibitem{Balle2011}
J.~Balle, A.~Stojanovic, and J.~Ohm, ``Models for static and dynamic texture
  synthesis in image and video compression,'' {\em IEEE Journal of Selected
  Topics in Signal Processing} {\bf 5}, 1353--1365  (2011).

\bibitem{fan2011}
F.~Zhang and D.~R. Bull, ``A parametric framework for video compression using
  region-based texture models,'' {\em IEEE Journal of Selected Topics in Signal
  Processing} {\bf 5}, 1378--1392  (2011).

\bibitem{karam2015}
K.~Naser, V.~Ricordel, and P.~L. Callet, ``Local texture synthesis: A static
  texture coding algorithm fully compatible with hevc,'' in {\em 2015
  International Conference on Systems, Signals and Image Processing (IWSSIP)},
  37--40  (2015).

\bibitem{fu2018}
C.~Fu, {\em et~al.}, ``Texture segmentation based video compression using
  convolutional neural networks,'' {\em Electronic Imaging}   (2018).
\newblock {Burlingame, CA, USA}.

\bibitem{av1-joshi2017}
U.~Joshi, {\em et~al.}, ``Novel inter and intra prediction tools under
  consideration for the emerging av1 video codec,'' {\em Proceedings of SPIE}
  {\bf 10396}, 10396 -- 10396 -- 13  (2017).

\bibitem{av1-liu2017}
Z.~Liu, {\em et~al.}, ``Adaptive multi-reference prediction using a symmetric
  framework,,'' {\em Electronic Imaging} {\bf 2017}(2), 65--72  (2017).

\bibitem{chen2018}
D.~Chen, {\em et~al.}, ``Multi-reference video coding using stillness
  detection,'' {\em Electronic Imaging}   (2018).
\newblock {Burlingame, CA, USA}.

\bibitem{AOM}
``Aom - alliance for open media.'' \url{http://www.aomedia.org/}.

\bibitem{global_motion}
S.~Parker, {\em et~al.}, ``Global and locally adaptive warped motion
  compensation in video compression,'' {\em Proceedings of the IEEE
  International Conference on Image Processing} , 275--279  (2017).
\newblock {Beijing, China}.

\bibitem{Simonyan2014}
K.~Simonyan and A.~Zisserman, ``Very deep convolutional networks for
  large-scale image recognition,'' {\em arXiv preprint} , arXiv:1409.1556
  (2014).

\bibitem{stex}
R.~Kwitt and P.~Meerwald, ``Stex: Salzburg texture image database.''
  \url{http://www.wavelab.at/sources/STex/}.

\bibitem{place365}
B.~Zhou, {\em et~al.}, ``Places: An image database for deep scene
  understanding,'' {\em arXiv preprint} , arXiv:1610.02055  (2016).

\bibitem{torch}
R.~Collobert, K.~Kavukcuoglu, and C.~Farabet, ``Torch7: A matlab-like
  environment for machine learning,'' {\em Proceedings of the BigLearn workshop
  at the Neural Information Processing Systems} , 1--6  (2011).
\newblock {Granada, Spain}.

\bibitem{chen1998}
C.~Chen, J.~Luo, and K.~J. Parker, ``Image segmentation via adaptive k-mean
  clustering and knowledge-based morphological operations with biomedical
  applications,'' {\em IEEE transactions on image processing} {\bf 7},
  1673--1683  (1998).

\bibitem{multilayer}
Z.~Liu, {\em et~al.}, ``Adaptive multireference prediction using a symmetric
  framework,'' {\em Proceedings of the IS\&T International Symposium on
  Electronic Imaging, Visual Information Processing and Communication VIII} ,
  {65--72(8)}  (2017).
\newblock {Burlingame, CA}.

\bibitem{FAST}
E.~Rosten and T.~Drummond, ``Fusing points and lines for high performance
  tracking,'' {\em Proceedings of the Tenth IEEE International Conference on
  Computer Vision (ICCV)}   (2005).
\newblock {Beijing, China}.

\bibitem{RANSAC}
M.~A. Fischler and R.~C.~B. Bolles, ``Random sample consensus: A paradigm for
  model fitting with applications to image analysis and automated
  cartography,'' {\em Commun. ACM} {\bf 24}, 381--395  (1981).

\end{thebibliography}
\bibliographystyle{spiejour}   



\end{spacing}
\end{document}